\documentclass[twoside,twocolumn,9pt]{article}
\usepackage{extsizes}
\usepackage[super,sort&compress,comma]{natbib} 
\usepackage[version=3]{mhchem}
\usepackage[left=1.5cm, right=1.5cm, top=1.785cm, bottom=2.0cm]{geometry}
\usepackage{balance}
\usepackage{mathptmx}
\usepackage{sectsty}
\usepackage{graphicx} 
\usepackage{lastpage}
\usepackage[format=plain,justification=justified,singlelinecheck=false,font={stretch=1.125,small,sf},labelfont=bf,labelsep=space]{caption}
\usepackage{float}
\usepackage{fancyhdr}
\usepackage{fnpos}
\usepackage[english]{babel}
\addto{\captionsenglish}{%
  
}
\usepackage{array}
\usepackage{droidsans}
\usepackage{charter}
\usepackage[T1]{fontenc}
\usepackage[usenames,dvipsnames]{xcolor}
\usepackage{setspace}
\usepackage[compact]{titlesec}
\usepackage{hyperref}
\usepackage{epstopdf}
\usepackage{soul}
\definecolor{cream}{RGB}{222,217,201}

\begin{document}

\pagestyle{fancy}
\thispagestyle{plain}
\fancypagestyle{plain}{
%%%HEADER%%%
\renewcommand{\headrulewidth}{0pt}
}
%%%END OF HEADER%%%

%%%PAGE SETUP - Please do not change any commands within this section%%%
\makeFNbottom
\makeatletter
\renewcommand\LARGE{\@setfontsize\LARGE{15pt}{17}}
\renewcommand\Large{\@setfontsize\Large{12pt}{14}}
\renewcommand\large{\@setfontsize\large{10pt}{12}}
\renewcommand\footnotesize{\@setfontsize\footnotesize{7pt}{10}}
\makeatother

\renewcommand{\thefootnote}{\fnsymbol{footnote}}
\renewcommand\footnoterule{\vspace*{1pt}% 
\color{cream}\hrule width 3.5in height 0.4pt \color{black}\vspace*{5pt}} 
\setcounter{secnumdepth}{5}

\makeatletter 
\renewcommand\@biblabel[1]{#1}            
\renewcommand\@makefntext[1]% 
{\noindent\makebox[0pt][r]{\@thefnmark\,}#1}
\makeatother 
\renewcommand{\figurename}{\small{Fig.}~}
\sectionfont{\sffamily\Large}
\subsectionfont{\normalsize}
\subsubsectionfont{\bf}
\setstretch{1.125} %In particular, please do not alter this line.
\setlength{\skip\footins}{0.8cm}
\setlength{\footnotesep}{0.25cm}
\setlength{\jot}{10pt}
\titlespacing*{\section}{0pt}{4pt}{4pt}
\titlespacing*{\subsection}{0pt}{15pt}{1pt}
%%%END OF PAGE SETUP%%%

%%%FOOTER%%%
\fancyfoot{}
\fancyfoot[LO,RE]{\vspace{-7.1pt}\includegraphics[height=9pt]{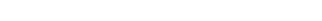}}
\fancyfoot[CO]{\vspace{-7.1pt}\hspace{13.2cm}\includegraphics{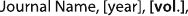}}
\fancyfoot[CE]{\vspace{-7.2pt}\hspace{-14.2cm}\includegraphics{head_foot/RF}}
\fancyfoot[RO]{\footnotesize{\sffamily{1--\pageref{LastPage} ~\textbar  \hspace{2pt}\thepage}}}
\fancyfoot[LE]{\footnotesize{\sffamily{\thepage~\textbar\hspace{3.45cm} 1--\pageref{LastPage}}}}
\fancyhead{}
\renewcommand{\headrulewidth}{0pt} 
\renewcommand{\footrulewidth}{0pt}
\setlength{\arrayrulewidth}{1pt}
\setlength{\columnsep}{6.5mm}
\setlength\bibsep{1pt}
%%%END OF FOOTER%%%

%%%FIGURE SETUP - please do not change any commands within this section%%%
\makeatletter 
\newlength{\figrulesep} 
\setlength{\figrulesep}{0.5\textfloatsep} 

\newcommand{\topfigrule}{\vspace*{-1pt}% 
\noindent{\color{cream}\rule[-\figrulesep]{\columnwidth}{1.5pt}} }

\newcommand{\botfigrule}{\vspace*{-2pt}% 
\noindent{\color{cream}\rule[\figrulesep]{\columnwidth}{1.5pt}} }

\newcommand{\dblfigrule}{\vspace*{-1pt}% 
\noindent{\color{cream}\rule[-\figrulesep]{\textwidth}{1.5pt}} }

\makeatother
%%%END OF FIGURE SETUP%%%

%%%TITLE, AUTHORS AND ABSTRACT%%%
\twocolumn[
  \begin{@twocolumnfalse}
{\includegraphics[height=30pt]{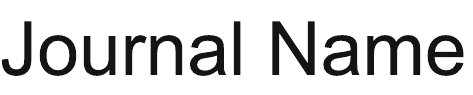}\hfill\raisebox{0pt}[0pt][0pt]{\includegraphics[height=55pt]{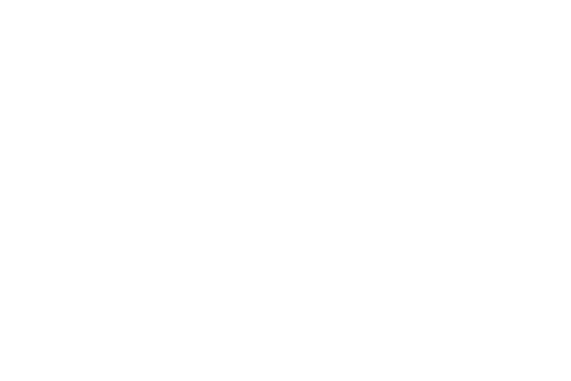}}\\[1ex]
\includegraphics[width=18.5cm]{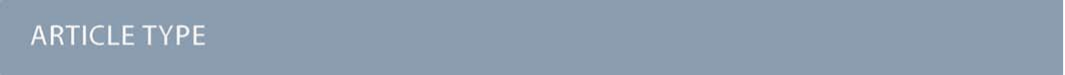}}\par
\vspace{1em}
\sffamily
\begin{tabular}{m{4.5cm} p{13.5cm} }

\includegraphics{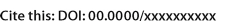} & \noindent\LARGE{\textbf{Novel 3D Pentagraphene Allotropes: Stability, Electronic, Mechanical, and Optical Properties}}
\\
\vspace{0.3cm} & \vspace{0.3cm}
\\

& \noindent\large{
 I. M. Félix{$^{a}$},
 B. Ipaves{$^{b}$},
 R. B. de Oliveira{$^{b}$},
 L. A. Ribeiro Junior{$^{c}$},
 L. S. Rocha\textit{$^{d}$}, 
 M. L. Pereira Junior{$^{e,\ast}$}, 
 D. S. Galv\~ao{$^{b}$} and 
 R. M. Tromer\textit{$^{c}$}
}
\\

\includegraphics{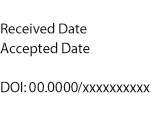} & \noindent\normalsize{Carbon-based materials have attracted great attention due to their exceptional structural diversity and wide-ranging applications. Recently, a new two-dimensional carbon allotrope, named pentagraphene (PG), was proposed. In this study, we proposed three novel three-dimensional (3D) PG allotropes, named 3D-PG-$\alpha$, -$\beta$, and -$\gamma$, engineered through biaxial strain and controlled compression of 2D PG layers. Comprehensive stability analyses, including phonon dispersion and \textit{ab initio} molecular dynamics simulations (AIMD), confirm their thermodynamic stability under room and high-temperature conditions. 3D-PG-$\alpha$ is the most stable, exhibiting a cohesive energy 0.5 eV/atom lower than the least stable structure, 3D-PG-$\gamma$. Electronic property characterization reveals semiconducting behavior for all structures, with indirect electronic band gaps ranging from 0.91 to 2.67 eV. The analyses of the mechanical properties showed significant anisotropy, with higher stiffness along the in-plane ($xy$-plane) direction. Optical properties highlight strong absorption along a wide range and a pronounced anisotropic response. Additionally, the absorption spectra exhibit activity in the visible region, and the refractive index and reflectivity indicate potential use in ultraviolet-blocking devices.}

\end{tabular}

 \end{@twocolumnfalse} \vspace{0.6cm}

  ]
%%%END OF TITLE, AUTHORS AND ABSTRACT%%%

%%%FONT SETUP - please do not change any commands within this section
\renewcommand*\rmdefault{bch}\normalfont\upshape
\rmfamily
\section*{}
\vspace{-1cm}

%%%FOOTNOTES%%%

\footnotetext{\textit{$^{a}$~
Center for Agri-food Science and Technology, Federal University of Campina Grande, Pombal, Paraíba, Brazil.}}
\footnotetext{\textit{$^{b}$~Department of Applied Physics and Center for Computational Engineering and Sciences, State University of Campinas, Campinas, São Paulo, Brazil.}}
\footnotetext{\textit{$^{c}$~University of Brasília, Institute of Physics, Brasília, Federal District, Brazil.}}
\footnotetext{\textit{$^{d}$~School of Engineering, MackGraphe, Mackenzie Presbyterian University, São Paulo, São Paulo, Brazil.}}
\footnotetext{\textit{$^{e}$~University of Bras\'{i}lia, College of Technology, Department of Electrical Engineering, Bras\'{i}lia, Federal District, Brazil.}}

%%%END OF FOOTNOTES%%%

%%%MAIN TEXT%%%%
\section*{Introduction}

Carbon allotropes, defined by the versatile arrangements of carbon atoms in several structural configurations, have become a cornerstone of materials science due to their exceptional properties and extensive range of potential applications \cite{hirsch2010era,georgakilas2015broad,ahmad2024carbon}. Among these allotropes, two-dimensional (2D) materials, particularly graphene \cite{novoselov2004electric,geim2007rise}, have revolutionized the field of carbon-based materials. Graphene’s extraordinary mechanical strength \cite{lee2008measurement,cao2020elastic}, electrical conductivity \cite{bolotin2008ultrahigh,lim2021measurements}, thermal conductivity \cite{balandin2008superior}, structural flexibility \cite{lee2015graphene}, and optical properties \cite{nair2008fine} have inspired the discovery and development of novel 2D structures. One such material is pentagraphene (PG), a 2D carbon allotrope proposed by Zhang \textit{et al.}, composed of pentagonal carbon rings, which exhibits mechanical robustness, thermal stability, and tunable electronic behavior \cite{zhang2015penta,nazir2022research}. These properties make PG an interesting structure for innovative material design \cite{quijano2017chiral,avramov2015translation}.

The advent of PG introduced a new paradigm in 2D materials, characterized by its unique pentagonal atomic arrangement and hybrid bonding nature \cite{nazir2022research}. Based on phonon dispersion and \textit{ab initio} molecular dynamics simulations (AIMD) \cite{zhang2015penta}, it has been demonstrated that PG can be thermodynamically stable up to high temperatures (up to 1000 K). Its buckled structure, with alternating sp\textsuperscript{2} and sp\textsuperscript{3} hybridized carbon atoms, distinguishes it from planar graphene and results in an intrinsic electronic band gap that ranges from 2.20 to 3.44 eV, depending on the computational method used (e.g., GGA, HSE06) \cite{zhang2015penta,li2019effect}. Moreover, PG exhibits a negative Poisson’s ratio, ultrahigh ideal strength, and piezoelectric properties, making it an attractive material for mechanical and electromechanical applications \cite{zhang2015penta,sun2016mechanical,guo2020tuning}. Despite its appealing properties, its experimental synthesis remains challenging, with current theoretical proposals focusing on the chemical exfoliation of precursors such as T12-carbon \cite{zhang2015penta}.

Recent advancements have further highlighted the potential of PG in various applications. Studies have demonstrated its suitability as a metal-free catalyst, a hydrogen storage material, and a channel material in field-effect transistors \cite{zhang2015penta,jia2018piezoelectric,guo2017all}. PG’s optical properties, including absorption in the ultraviolet range, and its tunable thermal conductivity also suggest promising roles in optoelectronics and thermoelectric devices \cite{einollahzadeh2016computing,alborznia2019buckling}. Furthermore, chemical functionalization and strain engineering have been shown to enhance its electronic and mechanical performance, enabling fine control over its bandgap and other properties \cite{zhang2017remarkable,shahrokhi2017tuning}. These advancements have spurred interest in PG-derived structures, such as nanoribbons, nanotubes, and multilayers, expanding the scope of research in pentagon-based materials \cite{zhang2015penta,zhang2017remarkable}.

Developing three-dimensional (3D) structures from 2D materials represents a transformative milestone in material science, unlocking unprecedented property spaces and functionalities. For PG, transitioning to 3D derivatives (3D-PG) enables the exploration of entirely new property regimes with tailored characteristics. However, achieving this transformation requires precise control over atomic rearrangements and interlayer interactions \cite{lee2019era,ipaves2019carbon,tromer2024transforming}. Recent advancements, including a strain-modulated approach \cite{tromer2024transforming}, have demonstrated that biaxial strain applied along the $xy$-plane combined with controlled compression of 2D layers can facilitate the formation of thermodynamically stable 3D allotropes. This strain-modulated methodology facilitates the synthesis of stable 3D structures. It allows for precise tuning of their mechanical and electronic properties, making it a versatile tool for advanced material design and future technological applications.

In this study, we employed the aforementioned strain-modulated methodology to predict three novel 3D-PG allotropes, designated as 3D-PG-$\alpha$, -$\beta$, and -$\gamma$. We have comprehensively analyzed their structural, electronic, mechanical, and optical properties. Structural stability analyses, including phonon dispersion and AIMD simulations, confirm that all three structures are thermodynamically stable at room and high-temperature conditions. Additionally, electronic property characterization reveals semiconducting behavior with indirect electronic band gaps ranging from 0.91 to 2.67 eV. Mechanical analyses indicate significant anisotropy, with higher stiffness observed along the $ xy$ plane. The optical behavior shows strong absorption and pronounced anisotropic responses, including activity in the visible spectrum and potential for ultraviolet-blocking applications. These findings highlight the transformative potential of strain-modulated methodologies.

\section*{Methodology}

The creation of the structural model for the proposed new 3D-PG structures from their 2D counterparts was carried out using the protocols described in \cite{tromer2024transforming}. The process began with three PG layers arranged with an initial interlayer spacing of 3 \r{A}. Biaxial strains were systematically applied along the $xy$-plane, followed by compression along the $z$-axis to induce interlayer bonding and form 3D structures. Stable configurations were achieved at specific biaxial strain levels of 2\%, 4\%, and 6\%, as identified through systematic exploration. Initial structural and energetic evaluations were performed using the Molecular Orbital PACkage (MOPAC2016) \cite{stewart1990mopac}, and the resulting structures were subsequently refined and analyzed using Density Functional Theory (DFT) simulations with the Spanish Initiative for Electronic Simulations with Thousands of Atoms (SIESTA) code \cite{soler2002siesta}.

The DFT simulations employed a double-$\zeta$ valence (DZV) basis set for localized atomic orbitals, with exchange-correlation effects treated using the Perdew-Burke-Ernzerhof (PBE) functional within the generalized gradient approximation (GGA) \cite{perdew1996generalized,ernzerhof1999assessment}. Norm-conserving Troullier-Martins pseudopotentials \cite{troullier1991efficient} accounted for core electrons. Convergence tests were conducted to ensure the accuracy of the calculations, including evaluations of $k$-point grid density, energy cutoff thresholds, and vacuum separation distances. Based on these tests, a kinetic energy cutoff of 300 Ry and a $5 \times 5 \times 5$ $k$-point grid were chosen for structural optimization, while a denser $15 \times 15 \times 15$ grid was employed for the electronic band structure and projected density of states (PDOS) calculations. During optimization, atomic positions and lattice vectors were fully optimized under periodic boundary conditions until atomic forces fell below 0.05 eV/\r{A}.

The stability of the proposed 3D-PG structures was evaluated through cohesive energy calculations and thermodynamic simulations. The cohesive energy ($E_\text{coh}$) was calculated using the following equation:
\begin{equation}
E_\text{coh} = \frac{1}{18}E_\text{3D-PG} - E_\text{C},
\end{equation}
where $E_\text{3D-PG}$ is the total energy of the unit cell containing 18 atoms, and $E_\text{C}$ is the energy of an isolated carbon atom. Phonon dispersion curves were computed at 0 K to confirm dynamical stability, and AIMD simulations were performed at 800 K in a canonical (NVT) ensemble to assess thermal stability.

The mechanical properties of the 3D structures were investigated by calculating Young's modulus ($E$). Stress-strain data were obtained by applying 1\% tensile and compressive strains separately along each crystallographic direction. Linear fits to the resulting stress-strain curves were used to extract $E^\text{T}$ and $E^\text{C}$, corresponding to tensile and compressive strains, respectively, enabling the evaluation of mechanical anisotropy in the structures.

The complex dielectric function $\epsilon(\omega)$ was calculated to characterize optical properties. The imaginary part, $\epsilon_2(\omega)$, was derived using Fermi’s golden rule \cite{tignon1995}, while the real part, $\epsilon_1(\omega)$, was obtained through Kramers-Kronig relations \cite{kronig1926,kramers1927,silveirinha2011examining}. These components were used to compute the absorption coefficient ($\alpha$), refractive index ($\eta$), and reflectivity ($R$) as follows:
\begin{equation}
\alpha (\omega) = \sqrt{2 \omega^2 \bigg(\sqrt{\epsilon_1^2(\omega) + \epsilon_2^2(\omega)} - \epsilon_1(\omega)\bigg)},
\end{equation}
\begin{equation}
\eta (\omega) = \frac{\sqrt{2}}{2} \bigg(\sqrt{\epsilon_1^2(\omega) + \epsilon_2^2(\omega)} + \epsilon_1(\omega)\bigg)^2,
\end{equation}
and
\begin{equation}
R(\omega) = \bigg(\frac{\sqrt{\epsilon_1(\omega) + i\epsilon_2(\omega)} - 1}{\sqrt{\epsilon_1(\omega) + i\epsilon_2(\omega)} + 1}\bigg)^2.
\end{equation}

\section*{Results and Discussion}

%%%%%%%%%%%%%%%%%%%%%%%%%%%
\begin{figure*}[htb!]
\centering
\includegraphics[width=0.7\linewidth]{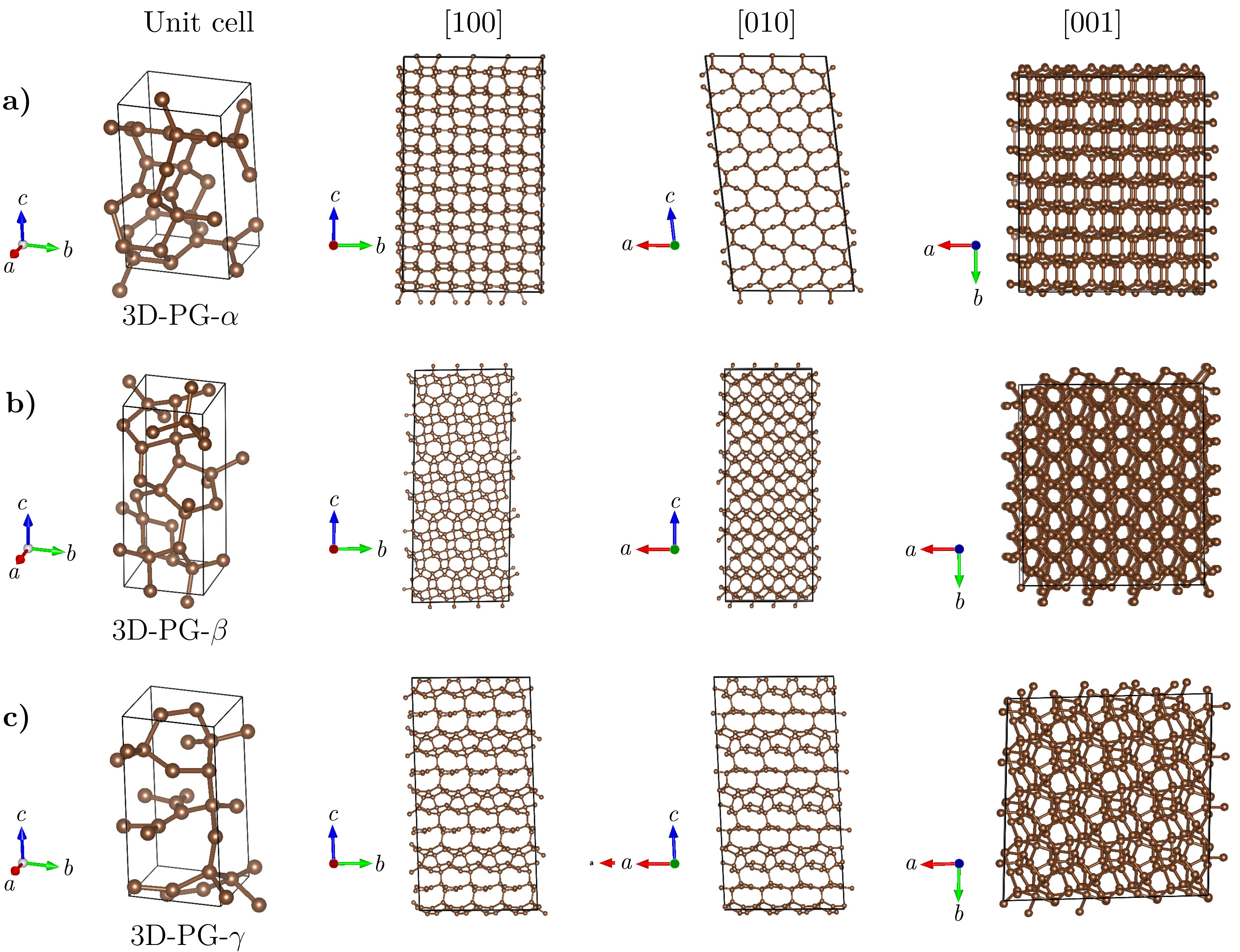}
\caption{Optimized 3D pentagraphene structures obtained for biaxial strain values of 2\%, 4\%, and 6\%, labeled as 3D-PG-$\alpha$ (a), 3D-PG-$\beta$ (b), and 3D-PG-$\gamma$ (c), respectively. On the left, the unit cell of each structure is displayed, followed by visualizations along the [100], [010], and [001] crystallographic directions.}
\label{fig:strucutures}
\end{figure*}
%%%%%%%%%%%%%%%%%%%%%%%%%%%

As mentioned above, the 3D structures investigated in this study were proposed using a recently developed methodology \cite{tromer2024transforming}, where layers of 2D systems are compressed to form 3D crystals. Using MOPAC16, we demonstrated that applying a biaxial strain along the $xy$-plane before compression along the $z$-direction produces distinct 3D structures, depending on the strain value. The process begins with three 2D PG layers separated by an initial interlayer spacing of 3 \r{A}. Biaxial strains ranging from 1\% to 10\% are applied, followed by sequential decreases in the $z$-axis lattice parameter values, in increments of 0.5 \r{A}, until an interlayer bonding occurs and a 3D crystal is formed. Among the configurations obtained, three distinct structures, corresponding to biaxial strain values of 2\% (3D-PG-$\alpha$), 4\% (3D-PG-$\beta$), and 6\% (3D-PG-$\gamma$), emerged as the most stable.

The optimized unit cells and their supercell representations along the [100], [010], and [001] crystallographic directions are presented in Figure \ref{fig:strucutures}. These visualizations highlight the structural distinctions between 3D-PG-$\alpha$, -$\beta$, and -$\gamma$, despite their common 2D precursor. The observed differences include variations in unit cell dimensions and anisotropic features along the $x$, $y$, and $z$-directions.

Table \ref{tab:structures} summarizes the structural parameters of the investigated 3D-PG systems, including lattice vector ($a$, $b$, and $c$) and angles ($\alpha$, $\beta$, and $\gamma$), carbon-carbon bond lengths ($R_\text{C-C}$), bond angles ($\theta_\text{C-C-C}$), and cohesive energy ($E_\text{coh}$). For 3D-PG-$\alpha$, bond lengths range from 1.35 \r{A} to 1.70 \r{A}, consistent with typical values for carbon-carbon bonds \cite{tromer2024transforming, enyashin2011graphene}. The range of bond angles, from $85^\circ$ to $118^\circ$, indicates a mixed hybridization state involving both $sp^2$ and $sp^3$ configurations. Similarly, for 3D-PG-$\beta$, bond lengths vary from 1.46 \r{A} to 1.68 \r{A}, with a wider range of bond angles (108$^\circ$ to 150$^\circ$), further emphasizing its structural uniqueness. In contrast, 3D-PG-$\gamma$ exhibits shorter bond lengths (1.33 \r{A} to 1.67 \r{A}) and more acute bond angles ($75^\circ$ to $111^\circ$), suggesting a predominantly strained configuration.

%%%%%%%%%%%%%%%%%%%%%%%%%%%
\begin{table*}[htb!]
\centering
\caption{Features of the optimized 3D-PG. Lattice vector constants ($a$, $b$, $c$) and angles ($\alpha$, $\beta$, $\gamma$), range of carbon-carbon bond lengths ($R_\text{C--C}$), bond angles ($\theta_\text{C--C--C}$), and cohesive energy ($E_\text{coh}$).}
\begin{tabular}{|c|c|c|c|c|c|c|c|c|c|}
\hline
Structure & $a$ (\r{A}) & $b$ (\r{A}) & $c$ (\r{A}) & $\alpha$ ($^\circ$) & $\beta$ ($^\circ$) & $\gamma$ ($^\circ$) & $R_\text{C--C}$ (\r{A}) & $\theta_\text{C--C--C}$ ($^\circ$) & $E_\text{coh}$ (eV/atom) \\
\hline
3D-PG-$\alpha$ & 3.75 & 4.31 & 7.35 & 90.00 & 83.00 & 90.00 & 1.35--1.70 & 85--118 & -8.87 \\
3D-PG-$\beta$ & 3.37 & 3.73 & 8.96 & 88.45 & 90.00 & 90.00 & 1.46--1.68 & 108--150 & -8.76 \\
3D-PG-$\gamma$ & 3.99 & 3.94 & 7.76 & 91.52 & 87.82 & 87.46 & 1.33--1.67 & 75--111 & -8.39 \\
\hline
\end{tabular}
\label{tab:structures}
\end{table*}
%%%%%%%%%%%%%%%%%%%%%%%%%%%

The calculated cohesive energies for 3D-PG-$\alpha$, -$\beta$, and -$\gamma$ are $-8.87$ eV/atom, $-8.76$ eV/atom, and $-8.39$ eV/atom, respectively. These results demonstrate a clear relationship between biaxial strain and structural stability: as the magnitude of the applied strain decreases, the resulting structures exhibit greater stability. The lowest cohesive energy of 3D-PG-$\alpha$ indicates it is the most stable structure, with a difference of approximately $0.5$ eV/atom compared to 3D-PG-$\gamma$. This trend suggests that reducing the biaxial strain facilitates the formation of energetically favorable configurations.

The analysis of structural anisotropy, as observed in Figure \ref{fig:strucutures}, reveals significant variations in the lattice dimensions among the three structures. For instance, the $c$ lattice parameter is the longest in 3D-PG-$\beta$ (8.96 \r{A}) compared to 3D-PG-$\alpha$ (7.35 \r{A}) and 3D-PG-$\gamma$ (7.76 \r{A}), reflecting the influence of strain-induced compression along the $z$-axis. This anisotropy aligns with the variations in bond angles and lengths, further supporting the role of biaxial strain in tailoring the mechanical and electronic properties of the 3D systems. 

%%%%%%%%%%%%%%%%%%%%%%%%%%%
\begin{figure*}[htb!]
\centering
\includegraphics[width=0.8\linewidth]{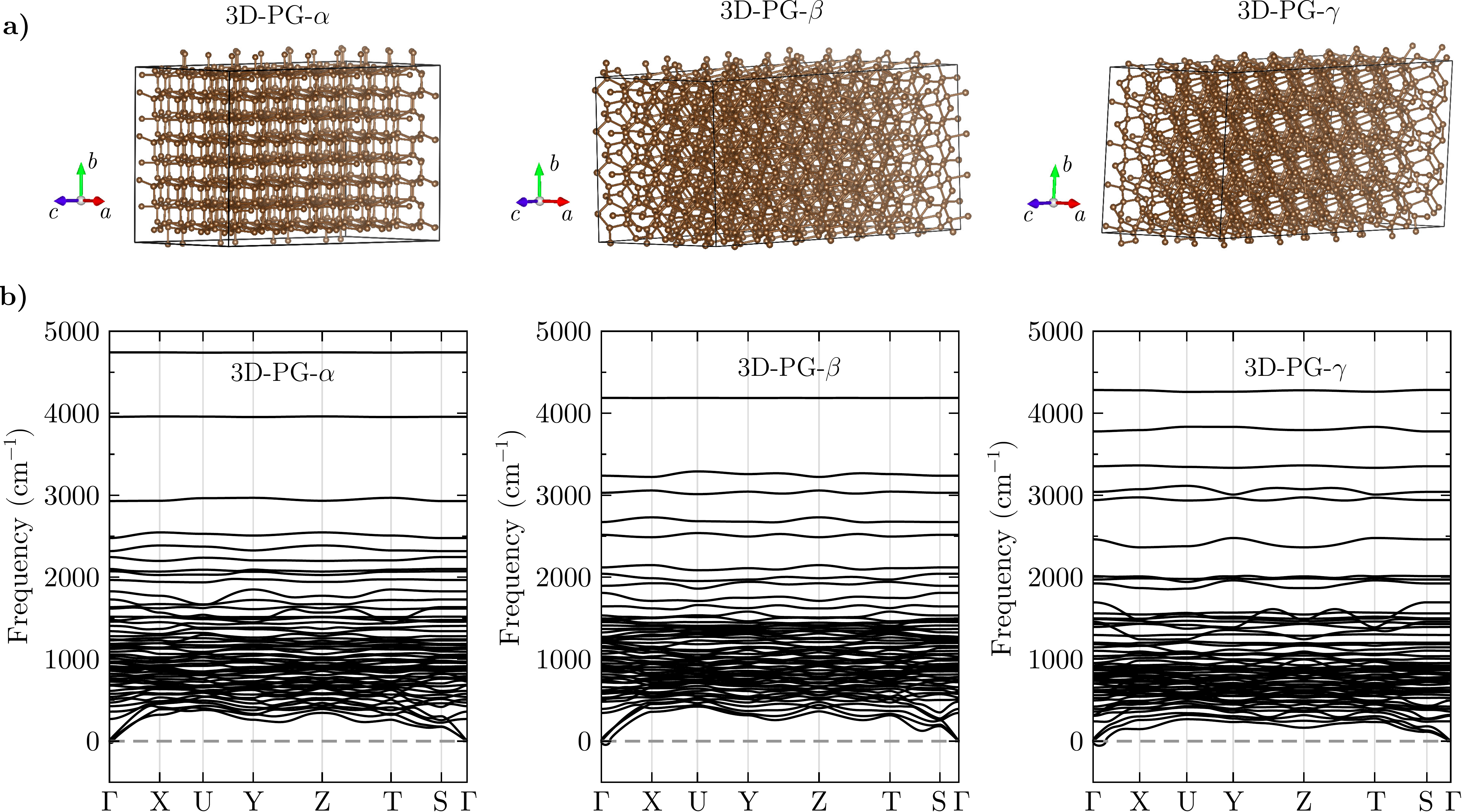}
\caption{Final configuration of the supercell obtained from AIMD simulations performed with an NVT ensemble at $T=800$ K for 3D-PG-$\alpha$, -$\beta$, and -$\gamma$ (a). Phonon dispersion spectra for the same structures along the symmetry paths (b).}
\label{fig:dispersion}
\end{figure*}
%%%%%%%%%%%%%%%%%%%%%%%%%%%

To further validate the robustness of these structures, their thermal stability under high-temperature conditions was investigated. Such validation is crucial for determining the feasibility of experimental synthesis processes. AIMD simulations were performed with an NVT ensemble at $T=800$ K for a total simulation time of 4 ps. As shown in Figure \ref{fig:dispersion}(a), which presents a snapshot of the supercell at the final step of the simulation, no significant distortions were observed when comparing the structures to their corresponding optimized geometries depicted in Figure \ref{fig:strucutures}. The absence of significant structural changes confirms that 3D-PG-$\alpha$, -$\beta$, and -$\gamma$ remain stable under high-temperature conditions.

To complement the thermal stability analyses, we calculated the phonon dispersion spectra of the three structures, which are presented in Figure \ref{fig:dispersion}(b). The frequency range extends up to approximately 5000 cm$^{-1}$ for 3D-PG-$\alpha$ and slightly above 4000 cm$^{-1}$ for 3D-PG-$\beta$ and -$\gamma$. In all cases, a high density of optical modes is observed between 50 cm$^{-1}$ and 1500 cm$^{-1}$, consistent with the vibrational activity typically associated with carbon-based materials. Importantly, no negative frequency values were observed in the acoustic modes along the symmetry paths, confirming the dynamical stability of the structures at $T=0$ K. These findings corroborate the results from AIMD simulations, providing strong evidence of the structural robustness of the 3D-PG structures.

The presence of high-frequency phonon modes, as indicated by the dispersion curves, suggests robust atomic bonding within the 3D frameworks, which aligns with prior studies on carbon-based allotropes \cite{chen2021tunable,felix20253d}. Such modes are often observed in three-dimensional materials stabilized under extreme conditions, such as high pressure \cite{trachenko2024upper,liu2024emergence,tsuppayakorn2022stabilizing}.

\begin{figure}[htb!]
\centering
\includegraphics[width=0.7\linewidth]{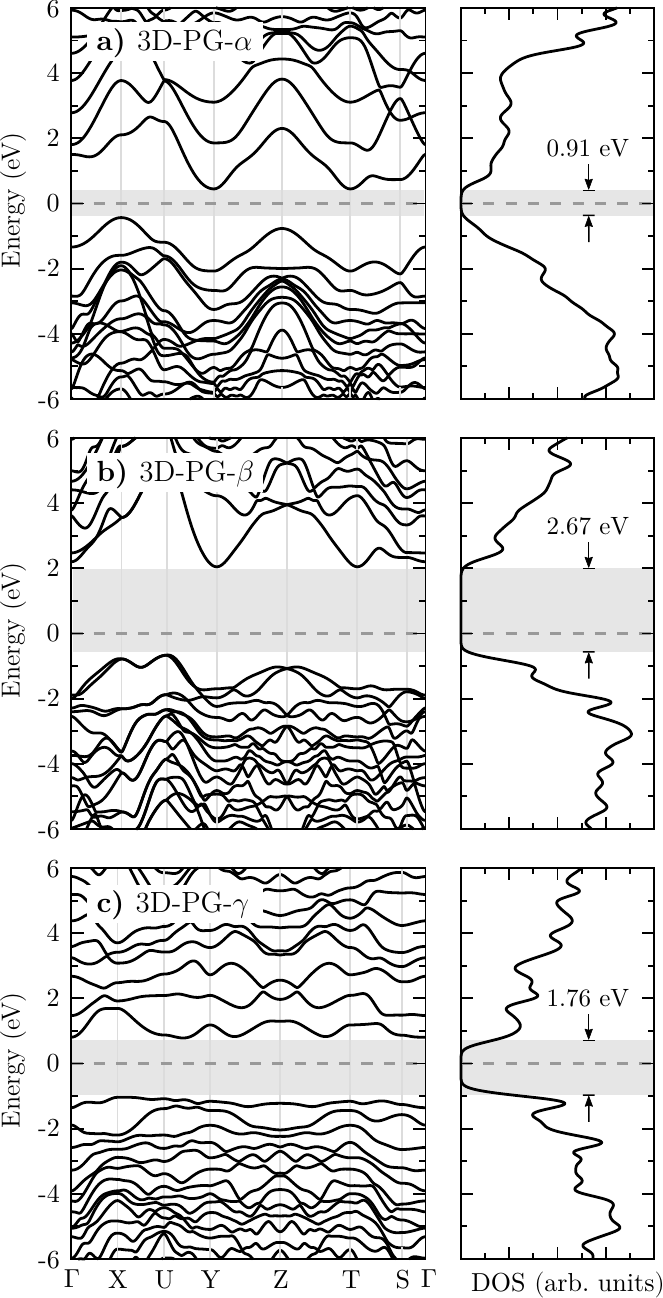}
\caption{Electronic band structure calculated along a high-symmetry path in the crystal and the corresponding density of states for structures 3D-PG-$\alpha$ (a), -$\beta$ (b), and -$\gamma$ (c).}
\label{fig:band-dos}
\end{figure}

Once we established the stability tests and obtained evidence that the structures were indeed stable, we performed band structure calculations to characterize the electronic nature of the structures. In Figure \ref{fig:band-dos}, we present the electronic band structure for each of the structures and the corresponding density of states for the valence states of the carbon atoms.

%%%%%%%%%%%%%%%%%%%%%%%%%%%
%%%%%%%%%%%%%%%%%%%%%%%%%%%

In general, we observed that all the structures are semiconductors with an indirect electronic band gap, with values of 0.91 eV, 2.67 eV, and 1.76 eV for structures 3D-PG-$\alpha$, $\beta$, and $\gamma$, respectively. The indirect electronic transitions occur between the $\text{X}\rightarrow\text{Y}$, $\text{U}\rightarrow\text{Y}$, and $\text{X}\rightarrow\Gamma$ points for structures 3D-PG-$\alpha$, -$\beta$, and -$\gamma$, respectively. The valence and conduction states of structures 3D-PG-$\alpha$ and -$\beta$ exhibit well-defined band curvature signatures, absent in structure 3D-PG-$\gamma$, where the valence bands are almost linear/flat. Although a more thorough future analysis of electronic transport is necessary, it is expected that 3D-PG-$\alpha$ and -$\beta$ will display better transport properties for both electrons and holes than 3D-PG-$\gamma$ since the effective masses tend to be smaller as a result of the well-defined concavities \cite{cheng2020high, ipaves2024tuning}. The effective masses were calculated for each structure, yielding values of $1.9\times 10^{-31}$ kg ($3.1\times 10^{-31}$ kg), $1.6\times 10^{-31}$ kg ($3.0\times 10^{-31}$ kg), and $1.9\times 10^{-31}$ kg ($17.8\times 10 ^{-31}$ kg) for electrons (holes) in structures 3D-PG-$\alpha$, -$\beta$, and -$\gamma$, respectively. As expected, 3D-PG-$\alpha$ and -$\beta$ exhibited similar practical mass values. However, for 3D-PG-$\gamma$, while the electron effective mass was comparable to the other two structures, the hole effective mass was higher, indicating poor hole transport properties.

Following the electronic characterization, the mechanical properties of the 3D-PG structures were analyzed to elucidate their potential for diverse applications. Young's modulus values were calculated as a measure of the stiffness of each structure. Small deformations corresponding to 1\% tensile and compressive strains were applied along each crystallographic direction ($x$, $y$, and $z$). The corresponding tensile ($E^\text{T}$) and compressive ($E^\text{C}$) Young's moduli were determined from the slope of a linear fit to the stress-strain curves. Table \ref{tab:stress} summarizes these values for each structure.

%%%%%%%%%%%%%%%%%%%%%%%%%%%
\begin{table*}[htb!]
\caption{Young's modulus for tension ($E^\text{T}$) and compression ($E^\text{C}$) along the $x$-, $y$-, and $z$-directions. Density values ($\rho$) are also included.}
\centering
\begin{tabular}{|c|c|c|c|c|c|c|c|}
\hline
Structure & $E_x^\text{T}$ (GPa) & $E_x^\text{C}$ (GPa)  & $E_y^\text{T}$ (GPa)  & $E_y^\text{C}$ (GPa)  & $E_z^\text{T}$ (GPa)  & $E_z^\text{C}$ (GPa) & $\rho$ ($10^{-27}$ kg/\r{A}$^3$) \\
\hline
3D-PG-$\alpha$ & 731.45 & 739.45 & 852.40 & 931.85 & 549.05 & 560.07 & 3.04 \\
3D-PG-$\beta$  & 685.53 & 698.99 & 942.56 & 983.68 & 670.45 & 701.96 & 3.18 \\
3D-PG-$\gamma$ & 632.80 & 690.89 & 756.47 & 803.29 & 353.01 & 376.61 & 2.94 \\
\hline
\end{tabular}
\label{tab:stress}
\end{table*}
%%%%%%%%%%%%%%%%%%%%%%%%%%%

Young's modulus values range from 353.0 GPa along the $z$-direction for 3D-PG-$\gamma$ to 983.7 GPa along the $y$-direction for 3D-PG-$\beta$. This variation correlates with the densities of the structures, with 3D-PG-$\beta$ exhibiting the highest density ($3.18 \times 10^{-27}$ kg/\r{A}$^3$) and 3D-PG-$\gamma$ the lowest ($2.94 \times 10^{-27}$ kg/\r{A}$^3$). The general trend in stiffness follows the order 3D-PG-$\beta$ > 3D-PG-$\alpha$ > 3D-PG-$\gamma$, indicating a direct relationship between density and mechanical strength.

These results also reveal significant anisotropies in the mechanical response of the 3D-PG structures. For all cases, compressive Young's modulus ($E^\text{C}$) exceeds tensile Young's modulus ($E^\text{T}$) along the same crystallographic direction. Additionally, both $E^\text{T}$ and $E^\text{C}$ are consistently lower along the $z$-direction compared to the $x$- and $y$-directions, except for 3D-PG-$\beta$, which exhibits relatively higher stiffness along the $z$-direction. This behavior is attributed to the formation process of the 3D crystals, where the 2D layers are initially bonded within the $xy$-plane before interlayer bonding along the $z$-direction. As a result, the interaction strength along the $z$-axis tends to be weaker than within the $xy$-plane.

The substantial Young's modulus values, notably exceeding 900 GPa in some cases, indicate that the 3D-PG structures possess stiffness comparable to that of other carbon-based materials, such as diamond and graphene. This high stiffness, combined with the observed anisotropy, underscores the potential of these structures for applications that require high mechanical strength and directional flexibility.

Finally, the optical properties of the 3D-PG structures were investigated by analyzing their interaction with polarized electromagnetic radiation in the photon energy range from 0 to 20 eV, encompassing the infrared, visible, and ultraviolet spectra. The polarization was considered along each structure's $x$-, $y$-, and $z$-directions. 

The absorption coefficient was initially analyzed, revealing no significant optical activity near the origin of the spectrum (infrared region). This behavior aligns with the semiconducting nature of the structures, where absorption occurs only for indirect electronic transitions corresponding to the optical gap. While the electronic and optical gaps are of similar magnitude, they are not identical, as evidenced by the absorption curves for all structures. For instance, for 3D-PG-$\alpha$, the absorption begins after 2.5 eV along the $x$-direction, slightly before 2.5 eV along the $y$-direction, and precisely at 2.5 eV in the $z$-direction. These variations indicate anisotropy in the optical gap, which can be attributed to direction-dependent electronic transitions. The absorption peaks vary in intensity across directions; for example, along the $y$-direction of 3D-PG-$\alpha$, the initial peaks exhibit a stronger optical activity compared to the other directions. 

Notably, all structures exhibit some degree of optical activity in the visible region. However, it is essential to acknowledge the limitations of using the GGA-PBE functional, which is well-known to underestimate electronic and optical gaps relative to more accurate methods, such as the hybrid HSE06 functional \cite{heyd2003hybrid, ipaves2024tuning}. A more precise calculation would shift the absorption spectrum toward the violet, potentially affecting the visibility of optical activity in the visible range. Maximum absorption in all structures occurs in the ultraviolet region, with photon energies ranging from 11 to 16 eV.
%%%%%%%%%%%%%%%%%%%%%%%%%%%
\begin{figure*}[htb!]
\centering
\includegraphics[width=0.8\linewidth]{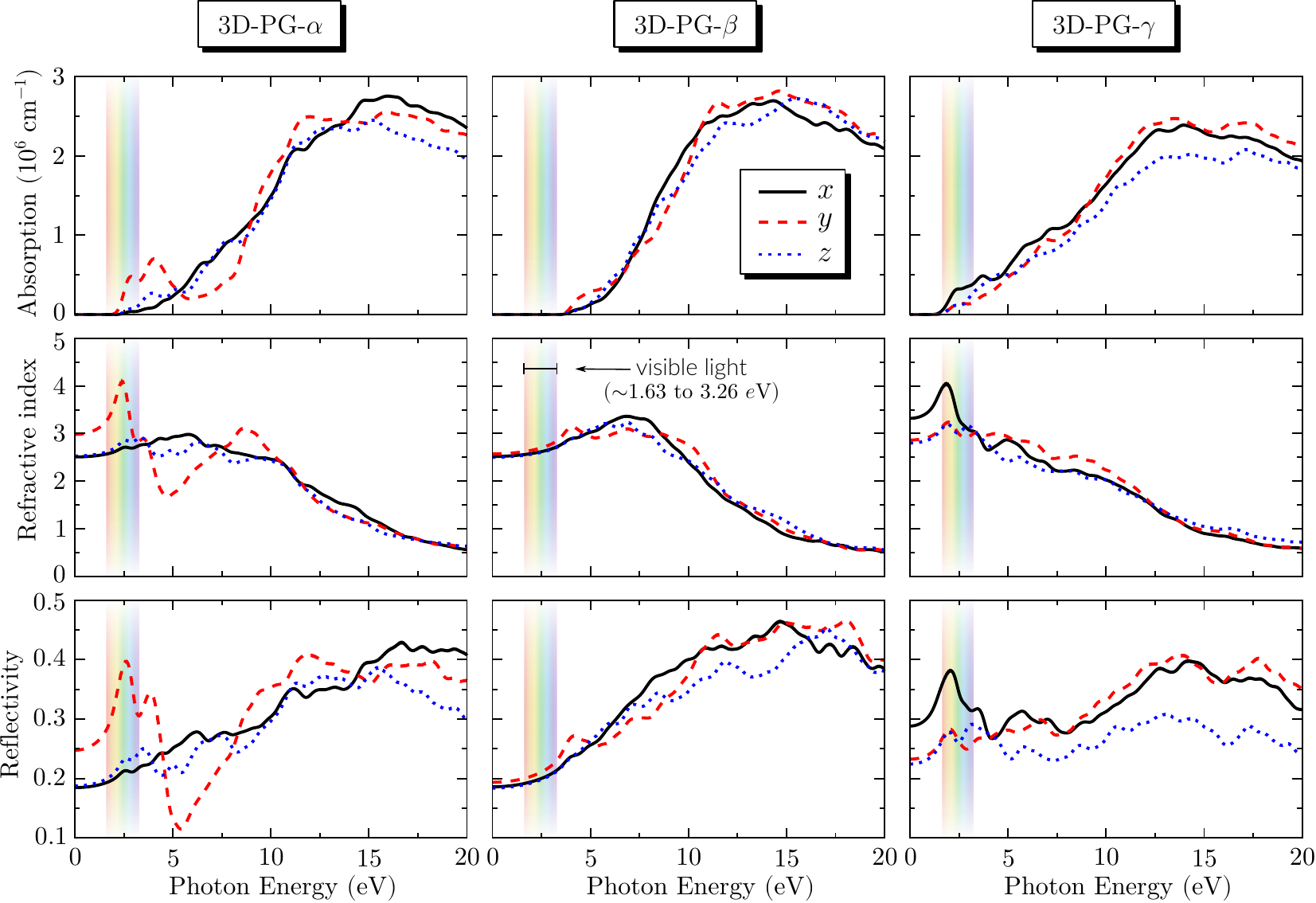}
\caption{Simulated optical properties of the 3D-PG structures: absorption coefficient, refractive index, and reflectivity as a function of photon energy, with polarization along the $x$-, $y$-, and $z$-directions.}
\label{fig:optical}
\end{figure*}
%%%%%%%%%%%%%%%%%%%%%%%%%%%

The refractive index was analyzed as a function of photon energy, revealing horizontal curves near the origin of the spectrum. These curves intersect the absorption axis, defining the static dielectric constants of the crystals \cite{srinivasu2012electronic}. For all structures, the refractive index remains relatively constant, up to approximately 10 eV, after which it decreases and approaches zero as the photon energy nears 20 eV. This behavior suggests that the refractive properties of the structures are more pronounced in the lower-energy regions of the spectrum, decreasing significantly in the ultraviolet range.

Reflectivity as a function of photon energy exhibits notable anisotropy, consistent with the trends in the absorption coefficient and refractive index. The reflectivity curves indicate that the structures reflect approximately 20-30\% of incident light in the red region, gradually increasing in the ultraviolet range. At the highest photon energy values, reflectivity approaches 50\%. Given that refraction decreases substantially in the ultraviolet region, it can be concluded that most incident light at these energy levels is reflected. This property highlights the potential of 3D-PG structures for applications as ultraviolet-blocking materials.

\section*{Conclusions}

In this study, we proposed and comprehensively characterized three novel 3D pentagraphene allotropes, designated as 3D-PG-$\alpha$, -$\beta$, and -$\gamma$. These structures were created using a strain-modulated methodology that combines the application of biaxial strain and controlled compression of 2D pentagraphene layers. The method produced thermodynamically stable 3D structures, confirmed by phonon dispersion calculations and AIMD simulations at 800 K. Among the three allotropes, 3D-PG-$\alpha$ was identified as the most stable, exhibiting a cohesive energy of approximately 0.5 eV/atom lower than 3D-PG-$\gamma$, the least stable structure.

Electronic property analyses revealed that all three allotropes exhibit semiconducting behavior with indirect electroninc band gaps ranging from 0.91 to 2.67 eV. Notably, 3D-PG-$\beta$ displayed the widest band gap, underscoring the potential for strain engineering to fine-tune electronic properties. This tunability makes these materials promising candidates for electronic and optoelectronic applications. 

Mechanical characterization highlighted significant anisotropy in Young’s modulus, with higher stiffness observed along the $xy$-plane compared to the $z$-direction. This mechanical behavior reflects the layered nature of the structures and their strain-induced interlayer bonding. The substantial stiffness values, comparable to those of other carbon-based materials, suggest potential use in applications requiring high mechanical strength and directional control.

Optical analyses revealed anisotropic responses for all three structures, with distinct absorption spectra that were influenced by the direction of polarization. While all structures exhibited absorption activity in the visible spectrum, the maximum absorption occurred in the ultraviolet range, with photon energies between 11 and 16 eV. The refractive index and reflectivity analyses highlight the potential for these materials to work as ultraviolet-blocking coatings or components in advanced photonic devices..

The findings presented in this work establish 3D-PG allotropes as a new class of carbon-based materials with tunable structural, electronic, mechanical, and optical properties. The proposed strain-modulated methodology demonstrates the feasibility of synthesizing these materials and provides a pathway for engineering their properties to tailor diverse technological applications. Future work should focus on experimental synthesis and real-world applications, further exploring the potential of these 3D carbon allotropes in next-generation nanomaterials and devices.

\section*{Author contributions}
I.M.F.: Formal analysis, Investigation, Visualization, Writing - Original Draft. 
B.I.: Data Curation, Investigation, Software, Writing - Original Draft. 
R.B.O.: Data Curation, Investigation, Software, Writing - Original Draft. 
M.L.P.J.: Formal analysis, Investigation, Resources, Funding acquisition, Writing - Review \& Editing. 
L.S.R.: Formal analysis, Investigation, Resources, Writing - Review \& Editing. 
D.S.G.: Formal analysis, Investigation, Resources, Funding acquisition, Writing - Review \& Editing. 
R.M.T.: Conceptualization, Methodology, Formal analysis, Investigation, Resources, Data Curation, Writing - Review \& Editing, Project administration.
All authors reviewed the manuscript. 

\section*{Conflicts of interest}
There are no conflicts to declare.

\section*{Data availability}
The data that support the findings of this study are available from the corresponding author, MLPJ, upon reasonable request.

\section*{Acknowledgements}

This work received partial support from Brazilian agencies CAPES, CNPq, and FAPDF.
M.L.P.J. acknowledges financial support from FAPDF (grant 00193-00001807/2023-16), CNPq (grants 444921/2024-9 and 308222/2025-3), and CAPES (grant 88887.005164/2024-00). Thanks are also extended to the National High-Performance Computing Center in São Paulo (CENAPAD-SP, State University of Campinas -- UNICAMP, project: proj960) and the High-Performance Computing Center (NACAD, Lobo Carneiro Supercomputer, Federal University of Rio de Janeiro -- UFRJ, project: a22002) for the computational support provided. B. I. acknowledges support from CNPq and São Paulo Research Foundation (FAPESP) process numbers 153733/2024-1 and 2024/11016-0. D. S. G. acknowledges the Center for Computing in Engineering and Sciences at Unicamp for financial support through the FAPESP/CEPID Grant \#2013/08293-7.

\bibliography{references} 

\providecommand*{\mcitethebibliography}{\thebibliography}
\csname @ifundefined\endcsname{endmcitethebibliography}
{\let\endmcitethebibliography\endthebibliography}{}
\begin{mcitethebibliography}{47}
\providecommand*{\natexlab}[1]{#1}
\providecommand*{\mciteSetBstSublistMode}[1]{}
\providecommand*{\mciteSetBstMaxWidthForm}[2]{}
\providecommand*{\mciteBstWouldAddEndPuncttrue}
  {\def\EndOfBibitem{\unskip.}}
\providecommand*{\mciteBstWouldAddEndPunctfalse}
  {\let\EndOfBibitem\relax}
\providecommand*{\mciteSetBstMidEndSepPunct}[3]{}
\providecommand*{\mciteSetBstSublistLabelBeginEnd}[3]{}
\providecommand*{\EndOfBibitem}{}
\mciteSetBstSublistMode{f}
\mciteSetBstMaxWidthForm{subitem}
{(\emph{\alph{mcitesubitemcount}})}
\mciteSetBstSublistLabelBeginEnd{\mcitemaxwidthsubitemform\space}
{\relax}{\relax}

\bibitem[Hirsch(2010)]{hirsch2010era}
A.~Hirsch, \emph{Nat. Mater.}, 2010, \textbf{9}, 868--871\relax
\mciteBstWouldAddEndPuncttrue
\mciteSetBstMidEndSepPunct{\mcitedefaultmidpunct}
{\mcitedefaultendpunct}{\mcitedefaultseppunct}\relax
\EndOfBibitem
\bibitem[Georgakilas \emph{et~al.}(2015)Georgakilas, Perman, Tucek, and Zboril]{georgakilas2015broad}
V.~Georgakilas, J.~A. Perman, J.~Tucek and R.~Zboril, \emph{Chem. Rev.}, 2015, \textbf{115}, 4744--4822\relax
\mciteBstWouldAddEndPuncttrue
\mciteSetBstMidEndSepPunct{\mcitedefaultmidpunct}
{\mcitedefaultendpunct}{\mcitedefaultseppunct}\relax
\EndOfBibitem
\bibitem[Ahmad \emph{et~al.}(2024)Ahmad, Mahmood, and Muhmood]{ahmad2024carbon}
F.~Ahmad, A.~Mahmood and T.~Muhmood, in \emph{{ACS} {Symposium} {Series}}, ed. T.~W. Quadri, C.~Verma, E.~E. Ebenso, M.~A. Quraishi and C.~M. Hussain, American Chemical Society, Washington, DC, 2024, vol. 1491, pp. 1--18\relax
\mciteBstWouldAddEndPuncttrue
\mciteSetBstMidEndSepPunct{\mcitedefaultmidpunct}
{\mcitedefaultendpunct}{\mcitedefaultseppunct}\relax
\EndOfBibitem
\bibitem[Novoselov \emph{et~al.}(2004)Novoselov, Geim, Morozov, Jiang, Zhang, Dubonos, Grigorieva, and Firsov]{novoselov2004electric}
K.~S. Novoselov, A.~K. Geim, S.~V. Morozov, D.-e. Jiang, Y.~Zhang, S.~V. Dubonos, I.~V. Grigorieva and A.~A. Firsov, \emph{Science}, 2004, \textbf{306}, 666--669\relax
\mciteBstWouldAddEndPuncttrue
\mciteSetBstMidEndSepPunct{\mcitedefaultmidpunct}
{\mcitedefaultendpunct}{\mcitedefaultseppunct}\relax
\EndOfBibitem
\bibitem[Geim and Novoselov(2007)]{geim2007rise}
A.~K. Geim and K.~S. Novoselov, \emph{Nat. Mater.}, 2007, \textbf{6}, 183--191\relax
\mciteBstWouldAddEndPuncttrue
\mciteSetBstMidEndSepPunct{\mcitedefaultmidpunct}
{\mcitedefaultendpunct}{\mcitedefaultseppunct}\relax
\EndOfBibitem
\bibitem[Lee \emph{et~al.}(2008)Lee, Wei, Kysar, and Hone]{lee2008measurement}
C.~Lee, X.~Wei, J.~W. Kysar and J.~Hone, \emph{Science}, 2008, \textbf{321}, 385--388\relax
\mciteBstWouldAddEndPuncttrue
\mciteSetBstMidEndSepPunct{\mcitedefaultmidpunct}
{\mcitedefaultendpunct}{\mcitedefaultseppunct}\relax
\EndOfBibitem
\bibitem[Cao \emph{et~al.}(2020)Cao, Feng, Han, Gao, Hue~Ly, Xu, and Lu]{cao2020elastic}
K.~Cao, S.~Feng, Y.~Han, L.~Gao, T.~Hue~Ly, Z.~Xu and Y.~Lu, \emph{Nat. Commun.}, 2020, \textbf{11}, 284\relax
\mciteBstWouldAddEndPuncttrue
\mciteSetBstMidEndSepPunct{\mcitedefaultmidpunct}
{\mcitedefaultendpunct}{\mcitedefaultseppunct}\relax
\EndOfBibitem
\bibitem[Bolotin \emph{et~al.}(2008)Bolotin, Sikes, Jiang, Klima, Fudenberg, Hone, Kim, and Stormer]{bolotin2008ultrahigh}
K.~I. Bolotin, K.~Sikes, Z.~Jiang, M.~Klima, G.~Fudenberg, J.~Hone, P.~Kim and H.~L. Stormer, \emph{Solid State Commun.}, 2008, \textbf{146}, 351--355\relax
\mciteBstWouldAddEndPuncttrue
\mciteSetBstMidEndSepPunct{\mcitedefaultmidpunct}
{\mcitedefaultendpunct}{\mcitedefaultseppunct}\relax
\EndOfBibitem
\bibitem[Lim \emph{et~al.}(2021)Lim, Park, Yamamoto, Lee, and Suk]{lim2021measurements}
S.~Lim, H.~Park, G.~Yamamoto, C.~Lee and J.~W. Suk, \emph{Nanomaterials}, 2021, \textbf{11}, 2575\relax
\mciteBstWouldAddEndPuncttrue
\mciteSetBstMidEndSepPunct{\mcitedefaultmidpunct}
{\mcitedefaultendpunct}{\mcitedefaultseppunct}\relax
\EndOfBibitem
\bibitem[Balandin \emph{et~al.}(2008)Balandin, Ghosh, Bao, Calizo, Teweldebrhan, Miao, and Lau]{balandin2008superior}
A.~A. Balandin, S.~Ghosh, W.~Bao, I.~Calizo, D.~Teweldebrhan, F.~Miao and C.~N. Lau, \emph{Nano Lett.}, 2008, \textbf{8}, 902--907\relax
\mciteBstWouldAddEndPuncttrue
\mciteSetBstMidEndSepPunct{\mcitedefaultmidpunct}
{\mcitedefaultendpunct}{\mcitedefaultseppunct}\relax
\EndOfBibitem
\bibitem[Lee \emph{et~al.}(2015)Lee, Kim, and Ahn]{lee2015graphene}
S.-M. Lee, J.-H. Kim and J.-H. Ahn, \emph{Mater. Today}, 2015, \textbf{18}, 336--344\relax
\mciteBstWouldAddEndPuncttrue
\mciteSetBstMidEndSepPunct{\mcitedefaultmidpunct}
{\mcitedefaultendpunct}{\mcitedefaultseppunct}\relax
\EndOfBibitem
\bibitem[Nair \emph{et~al.}(2008)Nair, Blake, Grigorenko, Novoselov, Booth, Stauber, Peres, and Geim]{nair2008fine}
R.~R. Nair, P.~Blake, A.~N. Grigorenko, K.~S. Novoselov, T.~J. Booth, T.~Stauber, N.~M. Peres and A.~K. Geim, \emph{Science}, 2008, \textbf{320}, 1308--1308\relax
\mciteBstWouldAddEndPuncttrue
\mciteSetBstMidEndSepPunct{\mcitedefaultmidpunct}
{\mcitedefaultendpunct}{\mcitedefaultseppunct}\relax
\EndOfBibitem
\bibitem[Zhang \emph{et~al.}(2015)Zhang, Zhou, Wang, Chen, Kawazoe, and Jena]{zhang2015penta}
S.~Zhang, J.~Zhou, Q.~Wang, X.~Chen, Y.~Kawazoe and P.~Jena, \emph{Proc. Natl. Acad. Sci. U. S. A.}, 2015, \textbf{112}, 2372--2377\relax
\mciteBstWouldAddEndPuncttrue
\mciteSetBstMidEndSepPunct{\mcitedefaultmidpunct}
{\mcitedefaultendpunct}{\mcitedefaultseppunct}\relax
\EndOfBibitem
\bibitem[Nazir \emph{et~al.}(2022)Nazir, Hassan, Shen, and Wang]{nazir2022research}
M.~A. Nazir, A.~Hassan, Y.~Shen and Q.~Wang, \emph{Nano Today}, 2022, \textbf{44}, 101501\relax
\mciteBstWouldAddEndPuncttrue
\mciteSetBstMidEndSepPunct{\mcitedefaultmidpunct}
{\mcitedefaultendpunct}{\mcitedefaultseppunct}\relax
\EndOfBibitem
\bibitem[Quijano-Briones \emph{et~al.}(2017)Quijano-Briones, Fern{\'a}ndez-Escamilla, and Tlahuice-Flores]{quijano2017chiral}
J.~Quijano-Briones, H.~Fern{\'a}ndez-Escamilla and A.~Tlahuice-Flores, \emph{Comput. Theor. Chem.}, 2017, \textbf{1108}, 70--75\relax
\mciteBstWouldAddEndPuncttrue
\mciteSetBstMidEndSepPunct{\mcitedefaultmidpunct}
{\mcitedefaultendpunct}{\mcitedefaultseppunct}\relax
\EndOfBibitem
\bibitem[Avramov \emph{et~al.}(2015)Avramov, Demin, Luo, Choi, Sorokin, Yakobson, and Chernozatonskii]{avramov2015translation}
P.~Avramov, V.~Demin, M.~Luo, C.~H. Choi, P.~B. Sorokin, B.~Yakobson and L.~Chernozatonskii, \emph{J. Phys. Chem. Lett.}, 2015, \textbf{6}, 4525--4531\relax
\mciteBstWouldAddEndPuncttrue
\mciteSetBstMidEndSepPunct{\mcitedefaultmidpunct}
{\mcitedefaultendpunct}{\mcitedefaultseppunct}\relax
\EndOfBibitem
\bibitem[Li \emph{et~al.}(2019)Li, Jin, Du, and Liu]{li2019effect}
L.~Li, K.~Jin, C.~Du and X.~Liu, \emph{RSC advances}, 2019, \textbf{9}, 8253--8261\relax
\mciteBstWouldAddEndPuncttrue
\mciteSetBstMidEndSepPunct{\mcitedefaultmidpunct}
{\mcitedefaultendpunct}{\mcitedefaultseppunct}\relax
\EndOfBibitem
\bibitem[Sun \emph{et~al.}(2016)Sun, Mukherjee, and Singh]{sun2016mechanical}
H.~Sun, S.~Mukherjee and C.~V. Singh, \emph{Physical Chemistry Chemical Physics}, 2016, \textbf{18}, 26736--26742\relax
\mciteBstWouldAddEndPuncttrue
\mciteSetBstMidEndSepPunct{\mcitedefaultmidpunct}
{\mcitedefaultendpunct}{\mcitedefaultseppunct}\relax
\EndOfBibitem
\bibitem[Guo and Wang(2020)]{guo2020tuning}
S.-D. Guo and S.-Q. Wang, \emph{Journal of Physics and Chemistry of Solids}, 2020, \textbf{140}, 109375\relax
\mciteBstWouldAddEndPuncttrue
\mciteSetBstMidEndSepPunct{\mcitedefaultmidpunct}
{\mcitedefaultendpunct}{\mcitedefaultseppunct}\relax
\EndOfBibitem
\bibitem[Jia \emph{et~al.}(2018)Jia, Mu, Li, Zhao, Wu, and Wang]{jia2018piezoelectric}
H.-J. Jia, H.-M. Mu, J.-P. Li, Y.-Z. Zhao, Y.-X. Wu and X.-C. Wang, \emph{Physical Chemistry Chemical Physics}, 2018, \textbf{20}, 26288--26296\relax
\mciteBstWouldAddEndPuncttrue
\mciteSetBstMidEndSepPunct{\mcitedefaultmidpunct}
{\mcitedefaultendpunct}{\mcitedefaultseppunct}\relax
\EndOfBibitem
\bibitem[Guo \emph{et~al.}(2017)Guo, Wang, and Wang]{guo2017all}
Y.~Guo, F.~Q. Wang and Q.~Wang, \emph{Applied Physics Letters}, 2017, \textbf{111}, 073503\relax
\mciteBstWouldAddEndPuncttrue
\mciteSetBstMidEndSepPunct{\mcitedefaultmidpunct}
{\mcitedefaultendpunct}{\mcitedefaultseppunct}\relax
\EndOfBibitem
\bibitem[Einollahzadeh \emph{et~al.}(2016)Einollahzadeh, Dariani, and Fazeli]{einollahzadeh2016computing}
H.~Einollahzadeh, R.~Dariani and S.~Fazeli, \emph{Solid State Communications}, 2016, \textbf{229}, 1--4\relax
\mciteBstWouldAddEndPuncttrue
\mciteSetBstMidEndSepPunct{\mcitedefaultmidpunct}
{\mcitedefaultendpunct}{\mcitedefaultseppunct}\relax
\EndOfBibitem
\bibitem[Alborznia \emph{et~al.}(2019)Alborznia, Naseri, and Fatahi]{alborznia2019buckling}
H.~Alborznia, M.~Naseri and N.~Fatahi, \emph{Superlattices and Microstructures}, 2019, \textbf{133}, 106217\relax
\mciteBstWouldAddEndPuncttrue
\mciteSetBstMidEndSepPunct{\mcitedefaultmidpunct}
{\mcitedefaultendpunct}{\mcitedefaultseppunct}\relax
\EndOfBibitem
\bibitem[Zhang \emph{et~al.}(2017)Zhang, Pei, Sha, Zhang, and Gao]{zhang2017remarkable}
Y.~Zhang, Q.~Pei, Z.~Sha, Y.~Zhang and H.~Gao, \emph{Nano Research}, 2017, \textbf{10}, 3865--3874\relax
\mciteBstWouldAddEndPuncttrue
\mciteSetBstMidEndSepPunct{\mcitedefaultmidpunct}
{\mcitedefaultendpunct}{\mcitedefaultseppunct}\relax
\EndOfBibitem
\bibitem[Shahrokhi(2017)]{shahrokhi2017tuning}
M.~Shahrokhi, \emph{Optik}, 2017, \textbf{136}, 205--214\relax
\mciteBstWouldAddEndPuncttrue
\mciteSetBstMidEndSepPunct{\mcitedefaultmidpunct}
{\mcitedefaultendpunct}{\mcitedefaultseppunct}\relax
\EndOfBibitem
\bibitem[Lee(2019)]{lee2019era}
H.-B.-R. Lee, \emph{The era of atomic crafting}, 2019\relax
\mciteBstWouldAddEndPuncttrue
\mciteSetBstMidEndSepPunct{\mcitedefaultmidpunct}
{\mcitedefaultendpunct}{\mcitedefaultseppunct}\relax
\EndOfBibitem
\bibitem[Ipaves \emph{et~al.}(2019)Ipaves, Justo, and Assali]{ipaves2019carbon}
B.~Ipaves, J.~F. Justo and L.~V. Assali, \emph{The Journal of Physical Chemistry C}, 2019, \textbf{123}, 23195--23204\relax
\mciteBstWouldAddEndPuncttrue
\mciteSetBstMidEndSepPunct{\mcitedefaultmidpunct}
{\mcitedefaultendpunct}{\mcitedefaultseppunct}\relax
\EndOfBibitem
\bibitem[Tromer \emph{et~al.}(2024)Tromer, Felix, Baughmann, Galvao, and Woellner]{tromer2024transforming}
R.~M. Tromer, L.~C. Felix, R.~H. Baughmann, D.~S. Galvao and C.~F. Woellner, \emph{J. Phys. Chem. A}, 2024, \textbf{128}, 7346--7352\relax
\mciteBstWouldAddEndPuncttrue
\mciteSetBstMidEndSepPunct{\mcitedefaultmidpunct}
{\mcitedefaultendpunct}{\mcitedefaultseppunct}\relax
\EndOfBibitem
\bibitem[Stewart(1990)]{stewart1990mopac}
J.~J. Stewart, \emph{J. Comput. Aided Mol. Des.}, 1990, \textbf{4}, 1--103\relax
\mciteBstWouldAddEndPuncttrue
\mciteSetBstMidEndSepPunct{\mcitedefaultmidpunct}
{\mcitedefaultendpunct}{\mcitedefaultseppunct}\relax
\EndOfBibitem
\bibitem[Soler \emph{et~al.}(2002)Soler, Artacho, Gale, Garc{\'\i}a, Junquera, Ordej{\'o}n, and S{\'a}nchez-Portal]{soler2002siesta}
J.~M. Soler, E.~Artacho, J.~D. Gale, A.~Garc{\'\i}a, J.~Junquera, P.~Ordej{\'o}n and D.~S{\'a}nchez-Portal, \emph{J. Phys.: Condens. Matter}, 2002, \textbf{14}, 2745\relax
\mciteBstWouldAddEndPuncttrue
\mciteSetBstMidEndSepPunct{\mcitedefaultmidpunct}
{\mcitedefaultendpunct}{\mcitedefaultseppunct}\relax
\EndOfBibitem
\bibitem[Perdew \emph{et~al.}(1996)Perdew, Burke, and Ernzerhof]{perdew1996generalized}
J.~P. Perdew, K.~Burke and M.~Ernzerhof, \emph{Physi. Rev. Lett.}, 1996, \textbf{77}, 3865\relax
\mciteBstWouldAddEndPuncttrue
\mciteSetBstMidEndSepPunct{\mcitedefaultmidpunct}
{\mcitedefaultendpunct}{\mcitedefaultseppunct}\relax
\EndOfBibitem
\bibitem[Ernzerhof and Scuseria(1999)]{ernzerhof1999assessment}
M.~Ernzerhof and G.~E. Scuseria, \emph{J. Chem. Phys.}, 1999, \textbf{110}, 5029--5036\relax
\mciteBstWouldAddEndPuncttrue
\mciteSetBstMidEndSepPunct{\mcitedefaultmidpunct}
{\mcitedefaultendpunct}{\mcitedefaultseppunct}\relax
\EndOfBibitem
\bibitem[Troullier and Martins(1991)]{troullier1991efficient}
N.~Troullier and J.~L. Martins, \emph{Phys. Rev. B}, 1991, \textbf{43}, 1993\relax
\mciteBstWouldAddEndPuncttrue
\mciteSetBstMidEndSepPunct{\mcitedefaultmidpunct}
{\mcitedefaultendpunct}{\mcitedefaultseppunct}\relax
\EndOfBibitem
\bibitem[Tignon \emph{et~al.}(1995)Tignon, Voisin, Delalande, Voos, Houdr{\'e}, Oesterle, and Stanley]{tignon1995}
J.~Tignon, P.~Voisin, C.~Delalande, M.~Voos, R.~Houdr{\'e}, U.~Oesterle and R.~Stanley, \emph{Phys. Rev. Lett.}, 1995, \textbf{74}, 3967\relax
\mciteBstWouldAddEndPuncttrue
\mciteSetBstMidEndSepPunct{\mcitedefaultmidpunct}
{\mcitedefaultendpunct}{\mcitedefaultseppunct}\relax
\EndOfBibitem
\bibitem[Kronig(1926)]{kronig1926}
R.~d.~L. Kronig, \emph{J. Opt. Soc. Am.}, 1926, \textbf{12}, 547--557\relax
\mciteBstWouldAddEndPuncttrue
\mciteSetBstMidEndSepPunct{\mcitedefaultmidpunct}
{\mcitedefaultendpunct}{\mcitedefaultseppunct}\relax
\EndOfBibitem
\bibitem[Kramers(1927)]{kramers1927}
H.~A. Kramers, Atti del Congresso Internationale dei Fisici, Como, Italy, 1927, pp. 1--13\relax
\mciteBstWouldAddEndPuncttrue
\mciteSetBstMidEndSepPunct{\mcitedefaultmidpunct}
{\mcitedefaultendpunct}{\mcitedefaultseppunct}\relax
\EndOfBibitem
\bibitem[Silveirinha(2011)]{silveirinha2011examining}
M.~G. Silveirinha, \emph{Phys. Rev. B}, 2011, \textbf{83}, 165119\relax
\mciteBstWouldAddEndPuncttrue
\mciteSetBstMidEndSepPunct{\mcitedefaultmidpunct}
{\mcitedefaultendpunct}{\mcitedefaultseppunct}\relax
\EndOfBibitem
\bibitem[Enyashin and Ivanovskii(2011)]{enyashin2011graphene}
A.~N. Enyashin and A.~L. Ivanovskii, \emph{physica status solidi (b)}, 2011, \textbf{248}, 1879--1883\relax
\mciteBstWouldAddEndPuncttrue
\mciteSetBstMidEndSepPunct{\mcitedefaultmidpunct}
{\mcitedefaultendpunct}{\mcitedefaultseppunct}\relax
\EndOfBibitem
\bibitem[Chen \emph{et~al.}(2021)Chen, Hu, Jia, Xie, and Liu]{chen2021tunable}
X.-K. Chen, X.-Y. Hu, P.~Jia, Z.-X. Xie and J.~Liu, \emph{Int. J. Mech. Sci.}, 2021, \textbf{206}, 106576\relax
\mciteBstWouldAddEndPuncttrue
\mciteSetBstMidEndSepPunct{\mcitedefaultmidpunct}
{\mcitedefaultendpunct}{\mcitedefaultseppunct}\relax
\EndOfBibitem
\bibitem[Felix \emph{et~al.}(2025)Felix, de~Oliveira, Ribeiro~Jr, Galv\~ao, Pereira~Jr, and Tromer]{felix20253d}
I.~M. Felix, R.~B. de~Oliveira, L.~A. Ribeiro~Jr, D.~S. Galv\~ao, M.~L. Pereira~Jr and R.~M. Tromer, \emph{ACS Omega}, 2025\relax
\mciteBstWouldAddEndPuncttrue
\mciteSetBstMidEndSepPunct{\mcitedefaultmidpunct}
{\mcitedefaultendpunct}{\mcitedefaultseppunct}\relax
\EndOfBibitem
\bibitem[Trachenko \emph{et~al.}(2024)Trachenko, Monserrat, Hutcheon, and Pickard]{trachenko2024upper}
K.~Trachenko, B.~Monserrat, M.~Hutcheon and C.~J. Pickard, \emph{ArXiv Preprint arXiv:2406.08129}, 2024\relax
\mciteBstWouldAddEndPuncttrue
\mciteSetBstMidEndSepPunct{\mcitedefaultmidpunct}
{\mcitedefaultendpunct}{\mcitedefaultseppunct}\relax
\EndOfBibitem
\bibitem[Liu \emph{et~al.}(2024)Liu, Li, Zurek, Zhuang, Liu, Yue, Guo, Zhang, Chi, Huang,\emph{et~al.}]{liu2024emergence}
Z.~Liu, J.~Li, E.~Zurek, Q.~Zhuang, Y.~Liu, J.~Yue, S.~Guo, A.~Zhang, Z.~Chi, X.~Huang \emph{et~al.}, \emph{Phys. Rev. B}, 2024, \textbf{109}, L180501\relax
\mciteBstWouldAddEndPuncttrue
\mciteSetBstMidEndSepPunct{\mcitedefaultmidpunct}
{\mcitedefaultendpunct}{\mcitedefaultseppunct}\relax
\EndOfBibitem
\bibitem[Tsuppayakorn-Aek \emph{et~al.}(2022)Tsuppayakorn-Aek, Phaisangittisakul, Ahuja, and Bovornratanaraks]{tsuppayakorn2022stabilizing}
P.~Tsuppayakorn-Aek, N.~Phaisangittisakul, R.~Ahuja and T.~Bovornratanaraks, \emph{Sci. Rep.}, 2022, \textbf{12}, 6700\relax
\mciteBstWouldAddEndPuncttrue
\mciteSetBstMidEndSepPunct{\mcitedefaultmidpunct}
{\mcitedefaultendpunct}{\mcitedefaultseppunct}\relax
\EndOfBibitem
\bibitem[Cheng \emph{et~al.}(2020)Cheng, Liu, and Liu]{cheng2020high}
T.~Cheng, Z.~Liu and Z.~Liu, \emph{Journal of Materials Chemistry C}, 2020, \textbf{8}, 13819--13826\relax
\mciteBstWouldAddEndPuncttrue
\mciteSetBstMidEndSepPunct{\mcitedefaultmidpunct}
{\mcitedefaultendpunct}{\mcitedefaultseppunct}\relax
\EndOfBibitem
\bibitem[Ipaves \emph{et~al.}(2024)Ipaves, Justo, Sanyal, and Assali]{ipaves2024tuning}
B.~Ipaves, J.~F. Justo, B.~Sanyal and L.~V. Assali, \emph{ACS Applied Electronic Materials}, 2024, \textbf{6}, 386--393\relax
\mciteBstWouldAddEndPuncttrue
\mciteSetBstMidEndSepPunct{\mcitedefaultmidpunct}
{\mcitedefaultendpunct}{\mcitedefaultseppunct}\relax
\EndOfBibitem
\bibitem[Heyd \emph{et~al.}(2003)Heyd, Scuseria, and Ernzerhof]{heyd2003hybrid}
J.~Heyd, G.~E. Scuseria and M.~Ernzerhof, \emph{The Journal of chemical physics}, 2003, \textbf{118}, 8207--8215\relax
\mciteBstWouldAddEndPuncttrue
\mciteSetBstMidEndSepPunct{\mcitedefaultmidpunct}
{\mcitedefaultendpunct}{\mcitedefaultseppunct}\relax
\EndOfBibitem
\bibitem[Srinivasu and Ghosh(2012)]{srinivasu2012electronic}
K.~Srinivasu and S.~K. Ghosh, \emph{The Journal of Physical Chemistry C}, 2012, \textbf{116}, 25015--25021\relax
\mciteBstWouldAddEndPuncttrue
\mciteSetBstMidEndSepPunct{\mcitedefaultmidpunct}
{\mcitedefaultendpunct}{\mcitedefaultseppunct}\relax
\EndOfBibitem
\end{mcitethebibliography}
\bibliographystyle{rsc} %the RSC's .bst file
\end{document}